\documentclass{aa}
\usepackage{graphicx}
\newcommand{\kmprs}  {\mbox{\rm km\,s$^{-1}$}}
\newcommand{\mA} {\mbox {m\AA}}
\newcommand{\feh} {\mbox{\rm [Fe/H]}}

\newcommand{\alphafe} {\mbox{\rm [$\alpha$/Fe]}}
\newcommand{\teff}  {\mbox{$T_{\rm eff}$}}
\newcommand{\logteff} {\mbox{${\rm log}\,T_{\rm eff}$}}
\newcommand{\logg}  {\mbox{{\rm log}\,$g$}}
\newcommand{\HII} {\ion{H}{ii}}

\newcommand{\CI} {\ion{C}{i}}
\newcommand{\OI} {\ion{O}{i}}

\newcommand{\FeII} {\ion{Fe}{ii}}

\newcommand{\NiI} {\ion{Ni}{i}}

\newcommand{\Mv} {\mbox{$M_V$}}

\newcommand{\lya} {Ly$\alpha$}
\def\ltsima{$\; \buildrel < \over \sim \;$}
\def\simlt{\lower.5ex\hbox{\ltsima}}
\def\gtsima{$\; \buildrel > \over \sim \;$}
\def\simgt{\lower.5ex\hbox{\gtsima}}
\begin{document}

\title{The evolution of the C/O ratio in metal-poor halo stars
\thanks{Based on observations collected at the European
Southern Observatory, Chile (ESO No. 67.D-0106)}}


\author{Chris~J.~Akerman \inst{1} 
\and Leticia~Carigi \inst{2}
\and Poul~E.~Nissen \inst{3}
\and Max~Pettini \inst{1} 
\and Martin~Asplund \inst{4}}


\institute{
Institute of Astronomy, University of Cambridge, Madingley Road, Cambridge, CB3 0HA, UK.
\email{cja@ast.cam.ac.uk, pettini@ast.cam.ac.uk}
\and Instituto de Astronom\'{\i}a,
Universidad Nacional Aut\'{o}noma de M\'{e}xico, A.P. 70-264, DF 04510, M\'{e}xico.
\email{carigi@astroscu.unam.mx}
\and Department of Physics and Astronomy, University of Aarhus, DK--8000
Aarhus C, Denmark.
\email{pen@phys.au.dk}
\and Research School of Astronomy and Astrophysics, Australian National University, 
Mount Stromlo Observatory, Cotter Road, Weston, ACT 2611, Australia.
\email{martin@mso.anu.edu.au}
}

\date{Received ..... / Accepted ......}

\abstract
{We report new measurements of carbon and oxygen
abundances in 34 F and G dwarf and subgiant stars belonging to
the halo population and spanning a range of
metallicity from [Fe/H]\,$= -0.7$ to $-3.2$\,.
The survey is based on observations of four
permitted lines of \CI\ near 9100\,\AA\ and 
the \OI\,$\lambda 7774$ triplet, all recorded
at high signal-to-noise ratios with the UVES
echelle spectrograph on the ESO VLT.
The line equivalent widths were analysed with
the 1D, LTE, MARCS model atmosphere code
to deduce C and O abundances; corrections due
to non-LTE and 3D effects are discussed.
When combined with similar published data for disk
stars, our results confirm the metallicity
dependence of the C/O ratio known from previous
stellar and interstellar studies: C/O drops
by a factor of $\sim 3-4$ as O/H decreases
from solar to $\sim 1/10$ solar. Analysed
within the context of standard models for the
chemical evolution of the solar vicinity,
this drop results from the metallicity dependence
of the C yields from massive stars with mass loss,
augmented by the delayed release of C from stars
of low and intermediate mass. The former is, however,
always the dominant factor. 
Our survey has also uncovered tentative evidence to
suggest that, as the oxygen abundance decreases 
below [O/H]\,$= -1$, [C/O] may not remain constant
at [C/O]\,$= -0.5$, as previously thought, but 
increase again, possibly approaching near-solar
values at the lowest metallicities
([O/H]\,$\simlt -3$). With the current dataset
this is no more than a $3 \sigma$ effect and it
may be due to metallicity-dependent non-LTE corrections
to the [C/O] ratio which have not been taken into account.
However, its potential importance as a window on the
nucleosynthesis by Population~III stars is
a strong incentive for future work,
both observational and theoretical,
to verify its reality. 
\keywords{Stars: abundances -- Galaxy: abundances -- Galaxy: evolution -- Galaxy: halo}}

\maketitle

\section{Introduction}

Since Eggen, Lynden-Bell, \& Sandage (\cite{eggen62}) showed that 
it is possible to study the formation history of the Galaxy using 
stellar abundances, such studies have become an integral part of 
our understanding of how galaxies evolve.  
During the Galaxy's evolution, nucleosynthesis by successive 
generations of stars proceeded along with dynamical processes.  
A fossil record of the state of the Galaxy at various 
epochs in its past is preserved in stars surviving from earlier 
times to the present; their chemical composition plays an important role
in identifying different stages in the formation of the 
Milky Way (Freeman \& Bland-Hawthorn 2002).
Furthermore, the relative abundances of pairs of elements---examined
as a function of metallicity---can provide clues to the main 
channels for their nucleosynthesis.

Prime examples are carbon and oxygen, the two
most abundant elements after hydrogen and helium  
and the first to be produced (after He) in the chain of 
stellar nucleosynthesis which enriched the universe from
its primordial mix of light elements to its composition today.
The current thinking is that oxygen is synthesised 
in massive stars and dispersed into 
the interstellar medium (ISM) by Type II supernovae.
Carbon, on the other hand, is produced during helium 
burning in stars of all masses,
but the dependence of its yield on stellar mass and 
initial composition is not well known.

Clues are provided by the behaviour of C/O vs. O/H.
Measurements of carbon and oxygen abundances in Galactic stars 
(Nissen \& Edvardsson 1992; Andersson \& Edvardsson \cite{andersson94}; 
Gustafsson et al. \cite{gustafsson99})
and \HII\ regions in spiral and irregular galaxies
(reviewed by Garnett 2003) show a decrease in C/O
by a factor of about three as the oxygen abundance 
decreases from solar to [O/H]\,$\simeq -1$.
This has been interpreted as resulting from the combination
of the metallicity dependence of the C yield from 
the winds of massive stars on the one hand,  
and the time delay in the release of C from intermediate and low mass stars
relative to the prompt enrichment of O from Type II supernovae on the other
(Henry, Edmunds, \& K\"{o}ppen 2000; 
Goswami \& Prantzos 2000; Carigi 2000; 
Chiappini, Romano, \& Matteucci 2003b).

The measurements of carbon and oxygen abundances
in disk stars cited above used 
the forbidden [\CI]~$\lambda 8727.13$ 
and [\OI]~$\lambda 6300.30$ lines.
However, in halo main sequence stars with [Fe/H]\,$< -1$ 
these lines become too weak to be measured reliably. 
Tomkin et al. (1992) showed that this low metallicity
regime can be probed with four stronger 
high excitation \CI\ lines 
near 9100\,\AA\ together with the oxygen triplet 
at $\lambda\lambda7771.94$, 7774.17, 7775.39\,.
The results of Tomkin et al. 
hint at interesting features in the behaviour of
the [C/O] ratio at low oxygen abundances, but the sample was 
small because of the practical difficulties 
of conducting high precision spectroscopy in the far red with
the instrumentation available ten years ago.
With the advent of large telescopes, equipped with
efficient echelle spectrographs and modern detectors, 
the time is now ripe to re-examine
the abundances of carbon and oxygen over the full range of
metallicities spanned by Galactic halo stars.
We address this topic in the present paper with 
new observations of 34 halo stars obtained with 
the Ultraviolet and Visual Echelle Spectrograph (UVES; Dekker et al. 2000)
on the European Southern Observatory {\it Very Large Telescope} ({\it VLT}).
The paper is arranged as follows. In Section 2 we give a brief 
description of the observations and data reduction. 
The properties of the 34 stars that make up our sample are
summarised in Section 3, while Section 4 presents the derivation
of the abundances of C and O, together with an assessment
of the likely errors. In Section 5 we discuss our findings in the
context of chemical evolution models of the solar vicinity.
Finally we summarise our main results and conclusions in Section 6.

\begin{figure}
\vspace{-3.25cm}
\hspace{-0.50cm}
\resizebox{10.03cm}{!}{\includegraphics{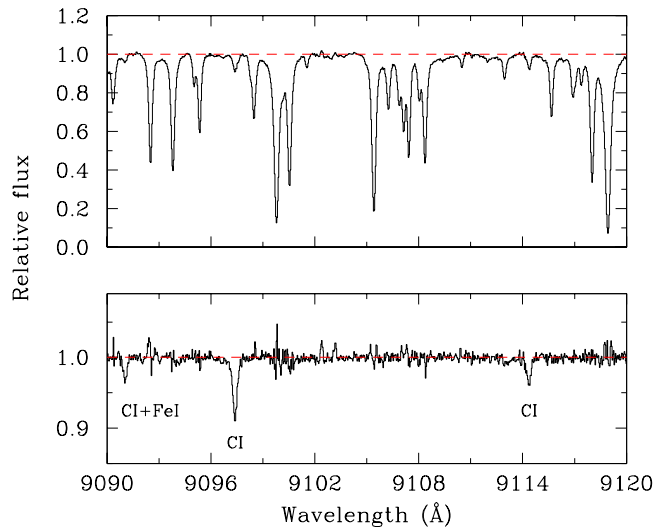}}
\vspace{-2.75cm}
\caption{The spectrum of CD$-35\degr14849$, 
before (top) and after (bottom) division by the
spectrum of a B-type star. The only remnant of the
telluric water vapour lines which mar this region 
of the optical spectrum is a higher level of noise
at the wavelengths of the H$_{2}$O lines.}
\label{fig:9090-9120}
\end{figure}

\section{Observations and data reduction}
\label{Obs}

Our sample consists of 34 F and G-type stars close to the main sequence
turn-off in the HR diagram. The stars, all with halo kinematics,
were selected mostly from the Str\"{o}mgren photometric catalogue
of Schuster \& Nissen (\cite{schuster88}) supplemented with a few
very metal-poor stars from Ryan, Norris, \& Beers (\cite{ryan99}). 
The selection criteria were: 
$5600 < \teff < 6500$\,K, $3.7 < \logg < 4.4$, and a smooth
distribution of metallicities from [Fe/H]\,=\,$-3.2$ to $-0.7$. 
The stellar visual magnitudes are between $V = 7.2$ and 11.5\,.

High resolution and signal-to-noise (S/N) spectra of the
34 stars were recorded with UVES in service mode between
March and June 2001. The total integration time was typically
one hour per star, divided between three 20 minute exposures
to facilitate the identification and removal of pixels
affected by cosmic-ray events. We used 
image slicer \#1, which gives a resolving power
$R \simeq 60\,000$, together with the DIC2 dichroic
which divides the spectrum into the blue and red 
arms with central wavelengths of 
4370\,\AA\ and 8600\,\AA\ respectively. 
All the \CI\ ($\lambda\lambda 9061.44, 9078.29$ and 
$\lambda\lambda 9094.83, 9111.81$) and \OI\ 
($\lambda\lambda 7771.94, 7774.17, 7775.39$)
lines of interest fall in the red portion of the spectrum.

\begin{figure*}
\vspace{-2.25cm}
\centerline{\resizebox{18cm}{!}{\includegraphics{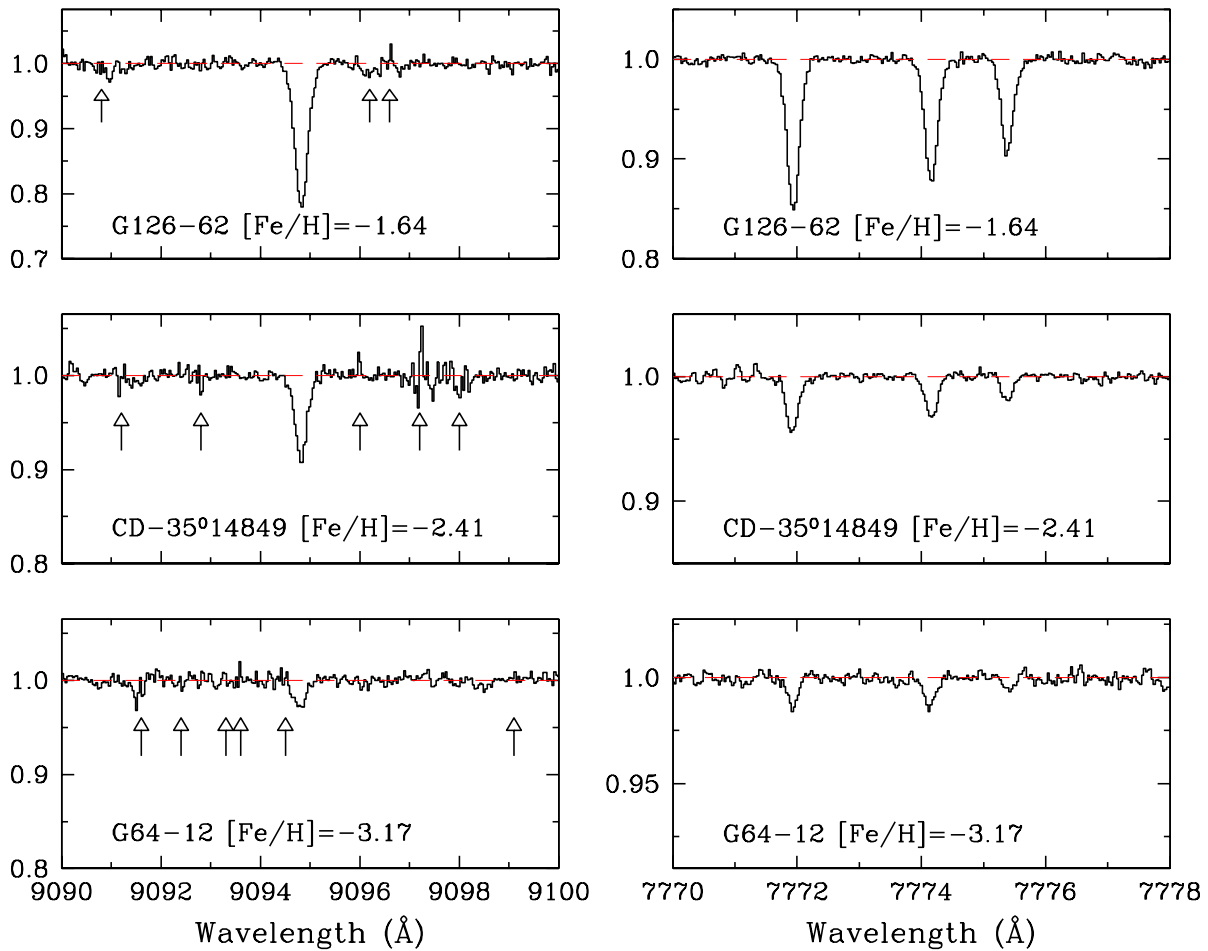}}}
\vspace{-8cm}
\caption{Sample portions of the UVES spectra, showing the
\CI~$\lambda 9094.83$ line ({\it left}) and 
the \OI\ $\lambda\lambda 7771.94, 7774.17, 7775.39$ triplet
({\it right}) in three stars with the metallicities indicated.
Arrows in the left-hand panels indicate where
telluric water vapour lines were divided out.
Note that in the star with the lowest metallicity in the present
sample, G64--12, the \CI\ and \OI\ lines can still 
be easily detected at [Fe/H]\,$= -3.17$\,.
}
\label{fig:CIOImulti}
\end{figure*}

The spectra were reduced with the standard echelle data reduction 
package in IRAF, following the usual steps of bias and scattered light 
subtraction, flat-fielding, order definition, wavelength calibration 
(by reference to the emission lines spectrum of the internal
Th-Ar lamp) and cosmic ray rejection 
(by comparing the three individual exposures of each star).
The far red portion of the optical spectrum where
the \CI\ lines are located is affected by the 
presence of several, strong, telluric water vapour lines.
They can, however, be removed effectively by division 
by the spectrum of a hot star (usually of B spectral type),
as shown in Figure 1.
This process also served to remove any residual fringing
which remained after flat-fielding.
At the shorter wavelengths of the \OI\ triplet telluric
absorption is not a problem, but the spectra of B stars
show some intrinsic absorption lines and could therefore
not be used to remove the residual fringing.
This was accomplished 
by fitting a high-order spline function to the spectra; 
with an amplitude of only a few percent and a period of $\sim 2$\,\AA, 
the fringing can be distinguished from the narrower stellar
absorption lines.
The final step in the data reduction consisted of correcting for
the stellar radial velocities. 
Representative portions of the spectra 
are reproduced in Figure \ref{fig:CIOImulti}.  
The typical signal-to-noise ratios per pixel near the spectral
features of interest are S/N\,$= 200-300$, measured
from the rms deviations from the continuum level.

\begin{figure}
\resizebox{\hsize}{!}{\includegraphics{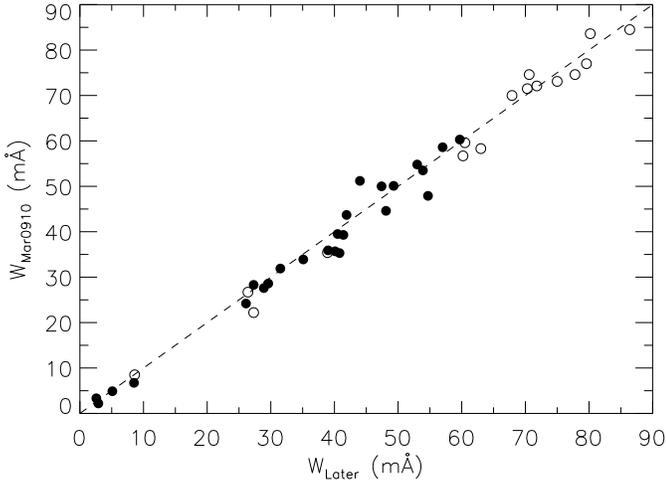}}
\caption{A comparison of the equivalent width measurements of 
\CI\ (open circles) and \OI\ (filled circles) lines between 
two separate observations of nine stars in our sample.  
The first set of observations was obtained on 9 and 10 March 2001;
their values are plotted on the $y$-axis. The second set was
acquired later on in the semester, between 12 March and 17 June
and their values are plotted on the $x$-axis.
The dotted line indicates the one-to-one relation.}
\label{fig:n1n2}
\end{figure}

\begin{figure}
\resizebox{\hsize}{!}{\includegraphics{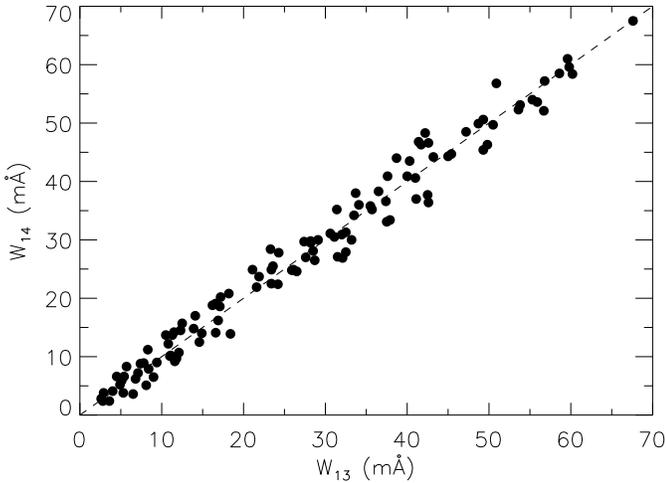}}
\caption{Comparison of \OI\ line equivalent width measurements 
from the 13th and 14th orders of the echelle spectrum of each star.  
The dotted line shows the one-to-one relation.}
\label{fig:1314}
\end{figure}

Equivalent widths ($W_{\lambda}$) of the \CI\ and \OI\ lines were measured 
by direct integration. We can estimate the errors in these
values empirically, by comparing (a) two independent spectra
available for nine stars which were observed twice for
operational reasons, and (b) the equivalent widths of the
\OI\ triplet lines which are recorded in two adjacent echelle orders
(orders 13 and 14).
The first comparison is shown in Figure \ref{fig:n1n2};
there is no systematic difference between the two sets of data and the standard 
deviations about the one-to-one relation are 3.0\,\mA\ and 2.9\,\mA\ 
for the \CI\ and \OI\ lines respectively. 
Similarly, Figure \ref{fig:1314} shows no systematic difference
between equivalent width measurements from the 13th and 14th echelle
orders and a standard deviation of 2.6\,\mA.
From these comparisons we estimate that the 
typical error applicable to our equivalent width
measurements is $\Delta W_{\lambda} \simeq 2$\,\mA\
($1/\sqrt 2$ of the standard deviations indicated by the above
comparisons between two sets of independent measurements).
This is confirmed by the comparison of our \OI\ equivalent
widths with the values published by Nissen et al. (2002)
for the seven stars which are in common between 
the two surveys; again we find no systematic offset 
and a standard deviation of 1.3\,m\AA.

\section{Stellar parameters}

\begin{table}
\caption[ ]{The derived values of effective temperature, surface gravity, 
microturbulence, absolute magnitude, mass and metal abundance.}
\label{table1}
\setlength{\tabcolsep}{0.10cm}
\begin{tabular}{lccccccr}
\noalign{\smallskip}
\hline
\noalign{\smallskip}
Star & $\teff$ & $\log g$ & $\xi_{\rm turb}$ & $M_V$ & Mass & $\feh$ \\ 
     & (K)     & (cgs)   & (km~s$^{-1}$) &      &($\cal{M}_{\odot}$)  \\ 
\noalign{\smallskip}
\hline
\noalign{\smallskip}
BD$-13\degr3442 $&    6500 &   4.16 &   1.5 &    4.03 &   0.75 &  $-$2.61 \\
CD$-30\degr18140$&    6272 &   4.13 &   1.5 &    4.06 &   0.75 &  $-$1.88 \\
CD$-35\degr14849$&    6125 &   4.11 &   1.5 &    4.24 &   0.70 &  $-$2.41 \\
CD$-42\degr14278$&    5812 &   4.25 &   1.5 &    4.83 &   0.70 &  $-$2.12 \\
HD\,103723   &        6040 &   4.26 &   1.3 &    4.36 &   0.87 &  $-$0.82 \\
HD\,105004   &        5919 &   4.36 &   1.2 &    4.78 &   0.83 &  $-$0.86 \\
HD\,106038   &        5919 &   4.30 &   1.2 &    4.84 &   0.70 &  $-$1.42 \\
HD\,108177   &        6034 &   4.25 &   1.5 &    4.61 &   0.70 &  $-$1.74 \\
HD\,110621   &        5989 &   3.99 &   1.5 &    3.95 &   0.75 &  $-$1.66 \\
HD\,121004   &        5595 &   4.31 &   1.0 &    5.02 &   0.76 &  $-$0.77 \\
HD\,140283   &        5690 &   3.69 &   1.5 &    3.47 &   0.75 &  $-$2.42 \\
HD\,146296   &        5671 &   4.17 &   1.2 &    4.57 &   0.80 &  $-$0.74 \\
HD\,148816   &        5823 &   4.14 &   1.2 &    4.21 &   0.89 &  $-$0.73 \\
HD\,160617   &        5931 &   3.77 &   1.5 &    3.35 &   0.82 &  $-$1.79 \\
HD\,179626   &        5699 &   3.92 &   1.2 &    3.92 &   0.80 &  $-$1.14 \\
HD\,181743   &        5863 &   4.32 &   1.5 &    4.96 &   0.70 &  $-$1.93 \\
HD\,188031   &        6054 &   4.03 &   1.5 &    4.02 &   0.72 &  $-$1.79 \\
HD\,193901   &        5672 &   4.38 &   1.0 &    5.31 &   0.66 &  $-$1.12 \\
HD\,194598   &        5906 &   4.25 &   1.3 &    4.63 &   0.75 &  $-$1.17 \\
HD\,215801   &        6005 &   3.81 &   1.5 &    3.47 &   0.76 &  $-$2.29 \\
LP\,815$-$43 &        6533 &   4.25 &   1.5 &    4.17 &   0.76 &  $-$2.67 \\
G\,011$-$044 &        5995 &   4.29 &   1.5 &    4.80 &   0.70 &  $-$2.09 \\
G\,013$-$009 &        6360 &   4.01 &   1.5 &    3.70 &   0.76 &  $-$2.27 \\
G\,016$-$013 &        5602 &   4.17 &   1.0 &    4.60 &   0.80 &  $-$0.76 \\
G\,018$-$039 &        5910 &   4.09 &   1.5 &    4.32 &   0.70 &  $-$1.52 \\
G\,020$-$008 &        5855 &   4.16 &   1.5 &    4.57 &   0.70 &  $-$2.28 \\
G\,024$-$003 &        5910 &   4.16 &   1.5 &    4.51 &   0.70 &  $-$1.67 \\
G\,029$-$023 &        5966 &   3.82 &   1.5 &    3.49 &   0.79 &  $-$1.80 \\
G\,053$-$041 &        5829 &   4.15 &   1.3 &    4.54 &   0.70 &  $-$1.34 \\
G\,064$-$012 &        6511 &   4.39 &   1.5 &    4.54 &   0.76 &  $-$3.17 \\
G\,064$-$037 &        6318 &   4.16 &   1.5 &    4.21 &   0.71 &  $-$3.12 \\
G\,066$-$030 &        6346 &   4.24 &   1.5 &    4.25 &   0.78 &  $-$1.52 \\
G\,126$-$062 &        5943 &   3.97 &   1.5 &    3.93 &   0.75 &  $-$1.64 \\
G\,186$-$026 &        6273 &   4.25 &   1.5 &    4.47 &   0.70 &  $-$2.62 \\
\hline
\end{tabular}
\end{table}

Table 1 lists relevant parameters of the stars observed.
A full description of the derivation of these parameters
is given in a companion paper (Nissen et al. 2003) which
reports measurements of the abundances of sulphur and zinc 
in the same stars. Here we summarise the most
important points.

Values of \teff\ were determined from the 
$b-y$ and $V-K$ colour indices using the
IRFM calibrations of Alonso et al. (\cite{alonso96})
as modified by Nissen et al. (\cite{nissen02}).  
The source of Str\"omgren $uvby$-$\beta$ photometry was
Schuster \& Nissen (\cite{schuster88}) for the large majority
of the stars, supplemented with unpublished photometry 
by Schuster et al. (\cite{schuster03}) for the remaining stars. 
Reddening corrections were included if $E$($b-y$) was
greater than 0.015 magnitudes.
The typical observational errors are 0.007\,mag in 
$b-y$ and 0.05\,mag in $V-K$, which correspond to an error 
of $\pm 50$~K in \teff\ in either case. Taking into account the 
uncertainty in the reddening, 
Nissen et al. (2003) estimate the 1\,$\sigma$
statistical error of $\teff$ to be around 70\,K.

Surface gravities were derived from the fundamental relation
\begin{eqnarray}
 \log \frac{g}{g_{\sun}} & = & \log \frac{\cal{M}}{\cal{M}_{\sun}}
 + 4 \log \frac{\teff}{T_{\rm eff,\sun}} + \\
    &   & 0.4 (M_{bol} - M_{bol,\sun})  \nonumber
\end{eqnarray}
where $\cal{M}$ is the mass of the star and $M_{bol}$ the absolute bolometric
magnitude.  
The absolute visual magnitude \Mv\ was determined from a new calibration
of the Str\"{o}mgren indices derived by Schuster et al.
(\cite{schuster03}) on the basis of Hipparcos parallaxes, and
also directly from the Hipparcos parallax (ESA \cite{esa97})
if available with an error $\sigma (\pi) / \pi < 0.3$. 
The bolometric correction was adopted from the work of
Alonso et al. (\cite{alonso95}), and the stellar mass was
obtained by interpolating in the \Mv -- \logteff\
diagram between the $\alpha$-element enhanced evolutionary tracks of
VandenBerg et al. (\cite{vandenberg00}). 
The combined effect of errors in \Mv, bolometric
correction, and stellar mass introduces an uncertainty 
of $\pm 0.15$\,dex in \logg .

Iron abundances were determined from 19 unblended \FeII\ lines
in the blue portion of the UVES spectra 
as described by Nissen et al. (2003).
In the more metal-poor stars ($\feh < -1.5$) the blue \FeII\ lines 
are so weak that the derived metallicity is practically independent of
the microturbulence. For such stars 
$\xi_{\rm micro} = 1.5$ \kmprs\ was assumed. 
For the more metal-rich stars,
$\xi_{\rm micro}$ was determined 
by requiring that the derived \feh\ values
should be independent of equivalent width.

\section{Carbon and oxygen abundances}

\subsection{Method}

The determination of abundances is based on $\alpha$-element enhanced
(\alphafe = +0.4, $\alpha$ = O, Ne, Mg, Si, S, Ca, and Ti) 
1D model atmospheres with the \teff , \logg , and \feh\ values
given in Table~1.
The models were computed with the MARCS code using updated continuous
opacities (Asplund et al. \cite{asplund97}) and including UV line blanketing
by millions of absorption lines.
Local thermodynamical equilibrium (LTE) is assumed 
both in constructing the models and in deriving abundances.
The Uppsala abundance analysis program, EQWIDTH,
was used to calculate theoretical equivalent widths
from the models. An elemental
abundance was then determined by requiring that the calculated
equivalent width match the observed value.

Table 2 lists the values of equivalent width used in the
abundance determinations.
Where no entry is given, it is either because the line 
is too strong or too weak to provide
a reliable abundance, or, in the case of \CI , 
is affected by residuals from the removal of strong telluric water vapour lines.
When more than one measurement of the same line is available
(for the reasons explained in Section 2), the values of $W_{\lambda}$ given
in Table~2 are averages of the individual measurements weighted
by their respective S/N ratios. 
For the \CI\ $\lambda\lambda 9061.44, 9078.29$ and 
$\lambda\lambda 9094.83, 9111.81$ lines we adopted 
values of $\log gf = -0.35, -0.58, +0.15, -0.30$
respectively, from the atomic spectra database
maintained by the National Institute of Standards and Technology.
For the \OI\ $\lambda\lambda 7771.94$, 7774.17, 7775.39
triplet we adopted $\log gf = +0.37, +0.22, 0.00$ respectively, 
from Wiese, Fuhr, \& Deterset (1996). 
The C and O abundances derived from each absorption line 
were finally averaged, giving to each line a weight
proportional to the strength ($gf$) of the transition.
The maximum deviation from the unweighted mean is less
than 0.05\,dex (usually in the most metal-poor stars).

\begin{table*}
\begin{center}
\caption[ ]{Measured equivalent widths of the four \CI\ 
lines near  9100\,\AA\ and the \OI~$\lambda 7774$ triplet, 
together with the derived oxygen and carbon abundances.  
The third and fifth columns of the right-hand panel
list abundances relative to the
1D, LTE, MARCS solar atmospheric values
$\log \epsilon \rm{(O)_{\odot} = 8.74}$ 
(Nissen et al. \cite{nissen02}) and
$\log \epsilon \rm{(C)_{\odot}} = 8.41$ 
(Allende Prieto, Lambert, \& Asplund \cite{prieto02}).}
\label{table2}
\setlength{\tabcolsep}{0.10cm}
\begin{tabular}{l|rrrr|rrr|ccccr}
\noalign{\smallskip}
\hline
\noalign{\smallskip}
Star &\multicolumn{4}{c|}{\CI\ $W_{\lambda}$ (m\AA)} &\multicolumn{3}{c|}{\OI\ $W_{\lambda}$ (m\AA)} & $\log \epsilon$(O)  & Non-LTE  & [O/H] & $\log \epsilon$(C) & [C/O] \\ 
     &$\lambda9061.4$ & $\lambda9078.3$ &$\lambda9094.8$ &$\lambda9111.8$ &$\lambda7771.9$ &$\lambda7774.2$ &$\lambda7775.4$ & (LTE) & correction$^{a}$ &(non-LTE) & (LTE) & (LTE)\\
\noalign{\smallskip}
\hline
\noalign{\smallskip}
BD$-13\degr3442 $&      &      &      & 10.4 &  9.8 &  6.0 &  5.1 & 6.85 & $-$0.10 & $-$1.99 & 6.13 & $-$0.39 \\ 
CD$-30\degr18140$& 24.6 & 17.5 & 50.8 & 29.3 & 25.4 & 23.0 & 14.5 & 7.55 & $-$0.13 & $-$1.32 & 6.72 & $-$0.50 \\
CD$-35\degr14849$&      &  9.4 & 26.8 & 12.5 & 10.6 &  8.1 &  4.1 & 7.12 & $-$0.11 & $-$1.73 & 6.40 & $-$0.39 \\
CD$-42\degr14278$&  9.9 &      & 34.1 & 15.8 & 11.5 &  8.2 &  5.1 & 7.42 & $-$0.13 & $-$1.45 & 6.62 & $-$0.47 \\ 
HD\,103723   &     72.6 & 58.5 &      & 74.1 & 57.8 & 49.7 & 37.9 & 8.34 & $-$0.20 & $-$0.60 & 7.63 & $-$0.38 \\
HD\,105004   &     69.0 & 60.7 &      & 78.3 & 45.9 & 37.2 & 28.8 & 8.26 & $-$0.19 & $-$0.67 & 7.75 & $-$0.18 \\   
HD\,106038   &     47.5 &      & 94.0 & 58.1 & 37.0 & 30.9 & 22.8 & 8.08 & $-$0.18 & $-$0.84 & 7.40 & $-$0.35 \\   
HD\,108177   &     32.9 & 18.5 & 56.2 & 30.4 & 31.6 & 23.3 & 14.4 & 7.80 & $-$0.15 & $-$1.09 & 6.94 & $-$0.53 \\  
HD\,110621   &     37.2 & 26.6 & 76.2 &      & 40.0 & 27.8 & 25.1 & 7.95 & $-$0.17 & $-$0.96 & 7.06 & $-$0.56 \\   
HD\,121004   &     70.9 & 60.1 &      &      & 53.7 & 48.7 & 34.5 & 8.71 & $-$0.23 & $-$0.26 & 8.05 & $-$0.33 \\   
HD\,140283   &      8.6 &  5.6 & 24.8 &  9.2 &  7.5 &  5.0 &  3.1 & 7.11 & $-$0.11 & $-$1.74 & 6.34 & $-$0.44 \\  
HD\,146296   &          &      &      & 81.9 & 53.9 & 42.8 & 31.7 & 8.53 & $-$0.22 & $-$0.43 & 7.96 & $-$0.24 \\
HD\,148816   &     90.6 & 78.0 &      &      & 67.6 & 59.3 & 44.7 & 8.65 & $-$0.22 & $-$0.31 & 8.02 & $-$0.30 \\   
HD\,160617   &     22.3 &      & 56.4 & 25.2 & 17.9 & 13.4 & 10.4 & 7.42 & $-$0.13 & $-$1.45 & 6.74 & $-$0.35 \\  
HD\,179626   &     66.0 & 45.0 &      & 67.3 & 54.4 & 50.1 & 33.3 & 8.46 & $-$0.22 & $-$0.50 & 7.62 & $-$0.51 \\   
HD\,181743   &     20.8 & 11.2 & 44.6 & 19.4 & 18.7 & 13.6 &  9.2 & 7.66 & $-$0.14 & $-$1.22 & 6.82 & $-$0.51 \\  
HD\,188031   &     28.8 & 17.7 & 60.0 & 35.0 & 30.8 & 23.0 & 17.9 & 7.75 & $-$0.15 & $-$1.14 & 6.87 & $-$0.55 \\   
HD\,193901   &     36.2 &      & 72.1 & 45.5 & 31.8 & 24.0 & 17.3 & 8.16 & $-$0.18 & $-$0.76 & 7.41 & $-$0.42 \\   
HD\,194598   &     50.4 & 44.1 &      & 63.6 & 42.6 & 35.6 & 28.6 & 8.19 & $-$0.19 & $-$0.74 & 7.48 & $-$0.38 \\   
HD\,215801   &     10.0 &  9.2 & 27.7 &      & 12.9 & 10.7 &  7.2 & 7.22 & $-$0.12 & $-$1.64 & 6.34 & $-$0.55 \\   
LP\,815$-$43 &          &      & 25.6 &  9.7 &  7.0 &  2.7 &      & 6.61 & $-$0.10 & $-$2.23 & 6.14 & $-$0.14 \\
G\,011$-$044 &     15.4 &      &      & 16.5 & 15.6 & 12.1 &  6.6 & 7.47 & $-$0.13 & $-$1.40 & 6.63 & $-$0.51 \\  
G\,013$-$009 &          &      & 42.1 & 20.1 & 15.4 & 12.5 &  8.4 & 7.15 & $-$0.11 & $-$1.70 & 6.47 & $-$0.35 \\   
G\,016$-$013 &     72.0 & 60.7 &      & 85.5 & 60.0 & 51.3 & 40.3 & 8.35 & $-$0.20 & $-$0.59 & 7.60 & $-$0.42 \\  
G\,018$-$039 &     46.9 & 30.8 &      & 56.8 & 41.8 & 36.9 & 25.7 & 8.12 & $-$0.18 & $-$0.80 & 7.29 & $-$0.50 \\  
G\,020$-$008 &     16.3 &  8.8 & 36.9 & 15.4 & 15.8 & 12.0 &  5.1 & 7.53 & $-$0.13 & $-$1.34 & 6.66 & $-$0.54 \\   
G\,024$-$003 &     19.3 &  7.9 & 48.9 & 19.5 & 18.5 & 16.8 & 11.3 & 7.62 & $-$0.14 & $-$1.26 & 6.71 & $-$0.58 \\   
G\,029$-$023 &     28.3 & 19.3 & 62.7 & 31.3 & 32.3 & 25.9 & 17.1 & 7.79 & $-$0.15 & $-$1.10 & 6.85 & $-$0.61 \\   
G\,053$-$041 &          &      & 65.4 & 30.9 &      & 21.8 & 14.1 & 7.86 & $-$0.16 & $-$1.04 & 7.07 & $-$0.46 \\
G\,064$-$012 &          &      &  8.7 &      &  3.1 &  3.2 &  1.2 & 6.44 & $-$0.10 & $-$2.40 & 5.68 & $-$0.43 \\ 
G\,064$-$037 &          &      &  9.5 &      &  3.0 &  2.5 &      & 6.44 & $-$0.10 & $-$2.40 & 5.72 & $-$0.39 \\
G\,066$-$030 &     36.6 & 25.6 & 75.4 & 35.4 & 46.4 & 38.1 & 28.2 & 7.93 & $-$0.13 & $-$0.97 & 6.93 & $-$0.67 \\   
G\,126$-$062 &     27.9 & 24.5 & 73.3 & 43.3 & 37.7 & 31.0 & 24.3 & 7.98 & $-$0.17 & $-$0.93 & 7.02 & $-$0.63 \\   
G\,186$-$026 &          &  4.2 & 19.2 &  9.1 &  4.5 &  3.8 &  3.4 & 6.71 & $-$0.10 & $-$2.13 & 6.14 & $-$0.24 \\
\hline
\end{tabular}
\end{center}
$^{a}$ $\log \epsilon({\rm O})_{\rm non-LTE} - \log \epsilon({\rm O})_{\rm LTE}$ 
correction factors calculated by interpolating
between the values given in Nissen et al. (2002).
\end{table*}

\subsection{Statistical abundance errors}

The main contributions to the random errors in the abundances arise
from the errors in the equivalent width measurements and from the 
uncertainties in the model atmosphere parameters. 
With typical values of $\Delta W_{\lambda} = 2$\,m\AA\ (Section 2),
the former dominate at low metallicities, while the latter
are proportionally more important in the more metal-rich stars.
Specifically, the $\pm 2$\,m\AA\ uncertainty in $W_{\lambda}$ translates
to abundance errors of only $\sim 5$\% in stars with
[O/H]\,$\simgt -1.5$, but as high as 25\% in the most metal-poor stars. 
(These values were obtained by 
appropriately varying the values of $W_{\lambda}$ 
supplied to the EQWIDTH program).
Abundance errors resulting from uncertainties in the
model atmospheres were estimated by changing
\teff\ of the stellar models by 70\,K, \logg\ by 0.15\,dex,
$\xi_{\rm turb}$ by $\pm 0.3$\,km~s$^{-1}$ and [Fe/H] by 0.2\,dex.
In general we found that these changes had relatively small
effects on the derived carbon and oxygen abundances.
The oxygen abundance changed by about 0.04\,dex
with either \teff\ or \logg, by $< 0.02$\,dex
with $\xi_{\rm turb}$,
and by $< 0.01$\,dex 
with [Fe/H].  
The C/O ratio is essentially
insensitive to all of these changes, since they affect the
C and O abundances similarly.

\subsection{Systematic abundance errors}

As in most abundance analyses, the uncertainties are likely to be 
dominated by systematic rather than statistical errors. 
Of particular concern are the
limitations and approximations inherent in our use of 1D hydrostatic
model atmospheres and the assumption of LTE for the line formation process.
The fact that the \CI\ and \OI\ lines used in this study
arise from levels with similar excitation potentials
(7.48 and 9.14\,eV respectively) may lead one to think that such
non-LTE and 3D effects may be comparable---and therefore may
have a relatively minor impact on the derived C/O ratios.
However, such an assumption 
clearly needs to be examined carefully.

\subsubsection{Non-LTE effects} 
The \OI\ $\lambda 7774$ triplet is known to
be susceptible to departures from LTE due to photon losses in the lines
(Sedlmayr 1974; Kiselman 1991). 
In terms of abundance, the impact of these non-LTE effects depends
on what is assumed for inelastic hydrogen collisions, which can help thermalise
the levels if the collision rates are sufficiently high. Unfortunately,
accurate laboratory or theoretical data for such collisions are scarce or, in
case of C and O, non-existent to our knowledge. Often, the classical Drawin (1968)
formula for H+H collisions is extrapolated to H collisions with other elements
to provide at least order-of-magnitude estimates. The available
data, however, suggest that the Drawin formula over-estimates the collisional
cross-sections by some three orders of magnitude (Fleck et al. 1991; Belyaev et al. 1999;
Barklem, Belyaev \& Asplund 2003; see also discussion in Kiselman 2001). 
Partly guided by these results, Kiselman (1991, 1993) and Nissen et al. (2002) 
ignored H collisions altogether and only treated
electron collisions. Depending on the stellar parameters,
they obtained typical non-LTE corrections of $-0.1$ to $-0.2$\,dex
to the abundances derived from the \OI\ triplet lines.
Here, we adopt the same approach and apply
corresponding non-LTE corrections to our O abundances
by interpolating between the values given in Nissen et al. (2002).
These corrections are listed in the second column of the 
right-hand panel of Table~2.

While oxygen has been the subject of several non-LTE investigations, 
the situation is significantly worse for carbon. 
In their study of metal-poor stars, 
Tomkin et al. (1992) carried out non-LTE
calculations for the same \CI\ and \OI\ lines used here.
These authors derived generally small correction factors: 
typically $\sim -0.03$\,dex
for \OI\ and $\sim -0.06$\,dex for \CI\ 
(although their sample included a few cases with much larger corrections),
and naturally concluded that the C/O ratio would be minimally affected 
by departures from LTE. Their calculations, however, assumed
very large values of the cross-sections
for H collisions; with smaller cross-sections
their conclusion may no longer hold.
In a preliminary analysis of the problem, we found 
that non-LTE effects on the \CI\ lines near 9100\,\AA\ 
may well be significantly larger than those for the \OI\ triplet,
although firm conclusions will only be reached when a proper
study is carried out.
It is difficult to estimate by how much the 
C/O ratios may have been overestimated under the assumption of LTE,
but factors of $\sim 0.2$\, dex 
cannot be excluded at this stage.
Furthermore, such non-LTE effects may have a metallicity
dependence in the sense of being more pronounced at lower metallicities,
thus complicating the interpretation of any trends of [C/O] vs [O/H].
Clearly, new non-LTE calculations for \CI\ are urgently needed to shed light on this
important issue; all we can do here is highlight the uncertainties
introduced by our current lack of knowledge of such effects 
and warn the reader that any conclusions reached below are dependent
upon this unresolved issue.

\begin{figure*}
\vspace{-2.25cm}
\centerline{\resizebox{17.5cm}{!}{\includegraphics[angle=270]{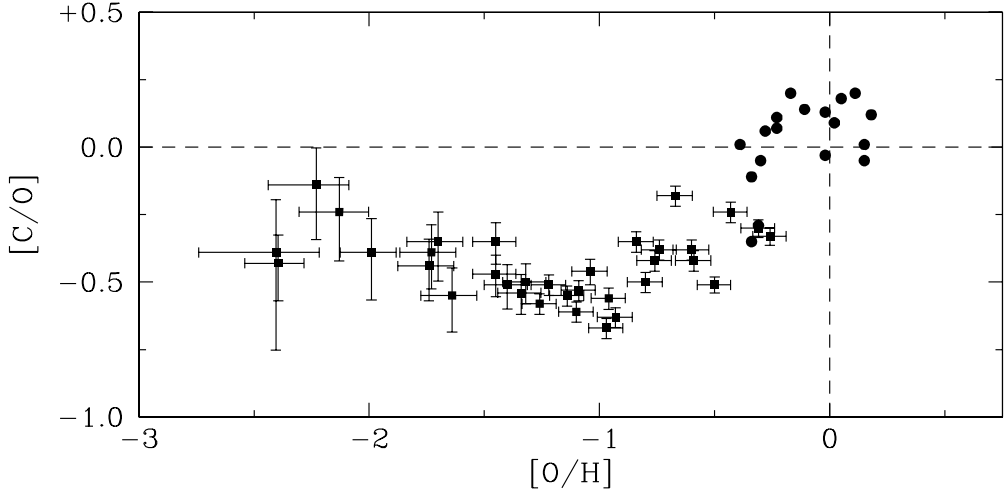}}}
\vspace{-3cm}
\caption{[C/O] vs [O/H] in Milky Way stars. 
Squares: data for halo stars from the VLT/UVES survey presented here.
Circles: data for disk stars from Nissen \& Edvardsson (1992);
Andersson \& Edvardsson (1994); and
Gustafsson et al. (1999).
The values of [O/H] have been corrected for non-LTE effects
(typically between 0 and $-0.2$\,dex---see Table~2 and text), 
but the values of [C/O] have not, because the non-LTE corrections
to the permitted \CI\ lines near 9100\,\AA\ used in this study
have not yet been reliably determined (see discussion in Section 4.3.1).
}
\label{fig:coplot}
\end{figure*}

\subsubsection{3D effects} 
Recently, a new generation of 3D hydrodynamical model atmospheres for
late-type stars has become available and applied to the derivation of element abundance 
(e.g. Stein \& Nordlund 1998; Asplund et al. 1999; 2000). Here we follow the same procedure
as in Asplund \& Garc\'{\i}a P{\'e}rez (2001) and Nissen et al. (2002, 2003) to investigate
the effects due to stellar granulation on the \CI\ and \OI\ lines using 3D models
with metallicities $-3.0 \leq {\rm [Fe/H]} \leq 0.0$. Under the assumption
of LTE, the \CI\ and \OI\ lines show small 3D effects and, importantly, of the same sign:
typically $<+0.05$\,dex. The similarity of the 3D effects is a consequence of the
similarly large line formation depths of these high-excitation lines. It is important to
remember, however, that significant non-LTE effects found in 1D model atmospheres,
as apparently is the case for the \OI\ and \CI\ lines, 
tend to be aggravated by the
large temperature inhomogeneities in the 3D models. 
Thus, ultimately it may be necessary to perform full
3D non-LTE calculations, particularly if
1D non-LTE studies of \CI\ indeed show large non-LTE effects.

\subsection{Results}

Having considered the statistical and systematic uncertainties
which may affect our abundance determinations, we now focus on 
our main results.
In the third and fifth columns of the right-hand panel
in Table~2 we have listed, respectively, the oxygen
and carbon abundances referred to the
solar values obtained from the 1D, LTE, MARCS 
model atmosphere analyses by Nissen et al. (\cite{nissen02})
[$\log \epsilon \rm{(O)_{\odot} = 8.74}$]
and Allende Prieto, Lambert, \& Asplund (\cite{prieto02})
[$\log \epsilon \rm{(C)_{\odot}} = 8.41$].
For comparison, the most recent determinations of the O and C
abundances using a full 3D solar model
are $\log \epsilon \rm{(O)_{\odot} = 8.66 \pm 0.05}$
(Asplund et al. 2003b)
and $\log \epsilon \rm{(C)_{\odot}} = 8.41 \pm 0.05$
(Asplund et al. 2003a);
thus, the values of [O/H] in Table~2 may be too low---and
the values of [C/O] too high---by 0.08\,dex.
In Figure~5 we plot [C/O] as a function of the
oxygen abundance (the latter corrected for non-LTE effects); 
the errors shown are the sum (in quadrature) of the different sources
of random error discussed in Section 4.2\,. 

We have included in Figure~5 earlier data for disk stars
by combining measurements 
of [C/H] by Andersson \& Edvardsson (\cite{andersson94}) and 
Gustafsson et al. (\cite{gustafsson99}), based on the
forbidden [\CI]\,$\lambda 8727.13$ line, with 
[O/H] determinations by Nissen \& Edvardsson (\cite{nissen92}) 
using [\OI]\,$\lambda 6300.30$ corrected for the \NiI\ blend
as discussed by Nissen et al. (2002). 
Both forbidden lines arise from low excitation
levels and are formed in the same layers of the stellar atmospheres. 
Hence, the derived C/O ratio is insensitive to errors 
in \teff\ and surface gravity. 
Corrections for 3D effects are expected to be
nearly the same for C and O. 
Non-LTE effects on the forbidden lines are also negligible;
see Kiselman (2001) for the case of [\OI]\,$\lambda 6300.30$ 
and St\"{u}renberg \& Holweger (1990) 
for the [\CI]\,$\lambda 8727.13$ line.
For these disk stars  
[O/H] and [C/H] were obtained by a 
differential analysis with respect to the Sun, using the same
forbidden lines in the solar and stellar spectra. 
Hence the adopted values of the absolute solar C and O abundances do 
not affect the derived [C/O] ratio in the same way as for the halo
stars (where, as discussed above, there may be an offset of 0.08\,dex).

Figure 5 confirms the known drop in [C/O] by a factor of 
$\sim 3-4$ as [O/H] decreases from solar to [O/H]\,$\simeq -1$ 
revealed by previous studies of carbon and oxygen abundances in 
Galactic stars (e.g. Tomkin et al 1992)
and in \HII\ regions of spiral and irregular
galaxies (e.g. Garnett 2003). The new data, however,
also suggest the intriguing possibility that the [C/O]
ratio may rise again at metallicities lower than
[O/H]\,$ = -1$.
Such a trend has not been recognised before and, as
discussed in Section 5 below, is not anticipated by 
most Galactic chemical evolution models---in all published
models [C/O] remains approximately flat for [O/H]\,$< -1$.
Its reality is far from secure, given the limited number
of stars with [O/H]\,$< -2$ in our sample and the large
errors, both random and systematic,
in the abundance determinations at such low
metallicities. In order to assess its 
statistical significance
in the present data set, we have performed a weighted
least-squares fit to the data in Figure~5 with 
[O/H]\,$< -1$, taking into account the random errors
in both [C/O] and [O/H].
The slope of the [C/O] vs. [O/H]
relation in this metallicity regime is $-0.21 \pm 0.06$, 
which is different from a flat relationship at the 
$3.5 \sigma$ level. If we consider only stars
with [O/H]\,$< -1.25$, the trend
is still significant at the $3 \sigma$ level
(the slope is $-0.25 \pm 0.08$). However, if
we limit ourselves to stars with [O/H]\,$< -1.5$,
the data are consistent with a flat relationship
(with slope of $-0.18 \pm 0.18$).
Note that in the first case the solar [C/O]
ratio would be recovered at [O/H]\,$ \simeq -3.7$,
{\it if} the trend continued beyond [O/H]\,$= -2.40$\,.

On the basis of such limited statistics it is not
possible to come to firm conclusions as to the 
behaviour of carbon at low oxygen abundances.
Rather, our results re-emphasise the importance of
continuing to survey these two important elements
in halo stars, particularly in the poorly explored
regime [O/H]\,$< -2$, where the traces left 
by the first generations of stars in the Galaxy
should be easiest to recognise.
Such surveys are well within the capabilities
of current astronomical instrumentation.

\section{Chemical evolution models}

\begin{table*}
\begin{center}
\caption[ ]{~Observational constraints and results of chemical 
evolution models of the solar neighbourhood.}
\label{}
\setlength{\tabcolsep}{0.12cm}
\begin{tabular}{lcc}
\noalign{\smallskip}
\hline
\noalign{\smallskip}
 Quantity & Observed Present Day ~ & ~ `Standard' Model  \\
          & Value$^{a}$            &    Value            \\
\hline
\noalign{\smallskip}
Total surface density $\sigma_{\rm gas+stars}$  ($M_{\odot}$ pc$^{-2}$)            & $51 \pm 6$      & 55   \\
\noalign{\smallskip}
Gas surface density   $\sigma_{\rm gas}$        ($M_{\odot}$ pc$^{-2}$)           & $13 \pm 3$      & 12.4 \\
\noalign{\smallskip}
Infall rate           $\dot{\sigma}_{\rm infall}$    ($M_{\odot}$ pc$^{-2}$ Gyr$^{-1}$)   & $0.9 \pm 0.6$   & 1.2  \\
\noalign{\smallskip}
Star formation rate                                                             &                 &      \\
surface density       $\Psi$  ($M_{\odot}$ pc$^{-2}$ Gyr$^{-1}$)                    & $3.5 \pm 1.5$   & 3.6  \\
\noalign{\smallskip}
Carbon abundance       $12 + \log {\rm(C/H)}$                                   & $8.39 \pm 0.06^{b}$ & 8.53 \\
\noalign{\smallskip}
Oxygen abundance       $12 + \log {\rm(O/H)}$                                   & $8.64 \pm 0.06^{b}$ & 8.99 \\
\noalign{\smallskip}
Solar carbon abundance $12 + \log {\rm(C/H)}_{\odot}$                           & $8.41 \pm 0.05^{c}$ & 8.35 \\
\noalign{\smallskip}
Solar oxygen abundance $12 + \log {\rm(O/H)}_{\odot}$                           & $8.66 \pm 0.05^{d}$ & 8.77 \\
\noalign{\smallskip}
\hline
\end{tabular}
\end{center}
$^{a}$ As compiled by Matteucci (2003).\\
$^{b}$ Orion nebula values (gas+dust) from Esteban et al. (1998, 2002).\\
$^{c}$ Allende Prieto, Lambert \& Asplund (2002); Asplund et al. (2003a).\\
$^{d}$ Allende Prieto, Lambert \& Asplund (2001); Asplund et al. (2003b).\\
\end{table*}

In this Section we attempt to interpret the pattern in the
abundances of carbon and oxygen shown in Figure~5 using chemical
evolution models for the solar neighbourhood. The basic principles
underlying this approach have been discussed extensively in the 
literature (see, for example, Matteucci 2003 for a recent review) 
so that here we only need to summarise the most important aspects. 
A number of (generally poorly known) parameters
describe the evolution of chemical elements 
(in this case C and O) through time, from the initial stages 
in the formation of the Milky Way to the present. The most
important ingredients are the history of star formation in the Galaxy,
the initial mass function (IMF) of successive stellar populations,
and the chemical yields of stars of different masses. 
With chemical evolution models we try to learn about 
all of these aspects from the observed abundances of elements,
while reproducing other available observational constraints.
Here we take the approach of adopting 
established ideas about the past history of star formation
in the solar neighbourhood and a conventional form of the IMF, 
and then consider what can be deduced about the yields of carbon
and oxygen as a function of metallicity from the behaviour
of [C/O] vs. [O/H]. 
The solar neighbourhood is defined as a cylinder, centred on the
Sun, of 1\,kpc radius and height. Within this volume, the 
observational constraints which must be met by any chemical 
evolution model are collected in Table 3. We now discuss the
model parameters in more detail.

\subsection{Star formation history}
Several previous works (e.g. Goswami \& Prantzos 2000;
Chiappini, Matteucci, \& Romano 2001;
Fenner \& Gibson 2003; Prantzos 2003a)
have successfully modelled
the evolution of the Milky Way with two infall episodes
(and no outflows);
the first gives rise to the halo over a short time scale,
while the disk is built more gradually during a second, protracted,
period of accretion. Thus, the total surface mass density
as a function of time is given by
\begin{equation}
\frac{d \sigma_{\rm gas+stars}}{dt} =
A e^{-t/\tau_{\rm halo}} + B e^{-(t-t_{\rm delay})/\tau_{\rm disk}}
\end{equation}
where the formation timescales 
$\tau_{\rm halo}$ and $\tau_{\rm disk}$ are taken to be 0.5 and 
6\,Gyr respectively, $t_{\rm delay} = 1$\,Gyr, and the constants
$A$ and $B$ are chosen so that the integral of equation (2)
over 13\,Gyr (taken to be the age of the Milky Way) matches the
present-day mass surface densities of the halo and disk 
(10 and 45\,$M_{\odot}$~pc$^{-2}$ respectively).
As discussed in the references given above, a
dual infall model with these parameters reproduces the
metallicity distributions of local K and G dwarf stars
and estimates of the present-day infall rate onto the disk.

In our models we assume that the infalling gas has primordial
composition. The star formation rate per unit area
$\Psi$, in units $M_{\odot}$~pc$^{-2}$~Gyr$^{-1}$, 
is proportional to the gas surface density,
$\Psi = \nu \times \sigma_{\rm gas}$. The constant
of proportionality $\nu$
measures the efficiency with which gas is converted into 
stars; as in previous work we assume a higher
efficiency during the halo formation than for the disk
with $\nu_{\rm halo} = 0.6$\,Gyr$^{-1}$ and 
$\nu_{\rm disk} = 0.3$\,Gyr$^{-1}$. These are the 
values which best reproduce the present-day values of 
$\sigma_{\rm gas}$ and $\Psi$ (Table 3), and the
transition from halo to disk at a metallicity
[O/H]\,$\simeq -0.5$\,.

\subsection{Initial mass function}

Not surprisingly, the choice of IMF can alter significantly
the outputs of chemical evolution models.
Here we adopt the three power-law approximation proposed
by Kroupa, Tout, \& Gilmore (1993, KTG) and apply it in the 
mass range $0.1 \leq m \leq 80 \,M_{\odot}$.
Other options are of course available. The KTG
formulation tends to be favoured in chemical evolution models
because of its relatively steep slope at the high mass end
[$\Phi(m) \propto m^{-\alpha}$ with $\alpha = 1.7$ for $m> 1\,M_{\odot}$]. 
Since oxygen is produced primarily by massive stars,
the value of $\alpha$ has a direct bearing on the oxygen 
abundance reached after a given fraction of the gas 
has been turned into stars.
It is a well known fact that IMFs with a flatter slope---such as 
that originally proposed by Salpeter (1955), with 
$\alpha = 1.35$ for stars with masses $m \geq 10 M_{\odot}$---tend 
to overproduce the present-day oxygen abundance 
when combined with standard stellar yields
(discussed in Section 5.3).
The problem is mitigated by the choice of the KTG IMF;
as can be seen from Table 3, our models come close
to reproducing the most recent estimates of (O/H)$_{\odot}$,
although they still overpredict somewhat the present-day
values measured in the Orion nebula (Esteban et al. 1998, 2002)
and in diffuse interstellar clouds (Oliveira et al. 2003 and
references therein).

\subsection{Stellar yields}

The stellar yields are another key ingredient 
of chemical evolution models.
Uncertainties can arise because 
even the most comprehensive sets of
published yields provide only a patchy coverage of the 
numerous combinations of stellar masses, metallicities,
mass-loss rates and other stellar properties, such as rotation
and binary fraction, which can affect 
the returned fraction of heavy elements.
For this reason, it is generally necessary to combine different 
sets of yields within any chemical evolution model.

In the present study we have adopted the yields of Meynet \& Maeder (2002)
and Meynet (2003, private communication)
for massive stars in the range $8~\leq~m~\leq~80$~$M_{\odot}$.
We use these massive star yields in preference to others (e.g. Woosley \& Weaver 1995)
because they take into account two important physical parameters, 
stellar rotation and mass loss through winds. 
Three sets of yields were calculated by Meynet \& Maeder (2002)
for initial metallicities (by mass) $Z = 0.02$, 0.004, and $1 \times 10^{-5}$
(respectively solar, 1/5 solar, and 1/2000 of solar); for each set
the following elements were considered: He, C, N, O, and the total
metal yield $Z$. Iron yields were not calculated explicitly, so that
for this element we use the yields by Woosley \& Weaver (1995).
In these authors' calculations, stars with masses of
$40 M_{\odot}$ produce no iron; thus we have assumed 
the iron yield to be zero for all stars with 
$m \geq 40 M_{\odot}$.

For stars in the mass range $0.8 \leq m \leq 8$\,$M_{\odot}$
(generally referred to as low- and intermediate-mass stars)
we use the comprehensive set of yields published by
van den Hoek \& Groenewegen (1997) for metallicities
0.04, 0.02, 0.008, 0.004, and 0.001 (from twice solar to 1/20 solar).
We kept these authors' mass loss scaling parameter $\eta_{\rm AGB}$ constant.
In our models 4\% of the stars with masses between 3 and 16\,$M_{\odot}$
are binary systems which explode as Type Ia supernovae with the yields calculated by
Thielemann, Nomoto, \& Hashimoto (1993). 

\begin{figure}
\vspace{-1.05cm}
\hspace{-0.750cm}
\resizebox{10.3cm}{!}{\includegraphics{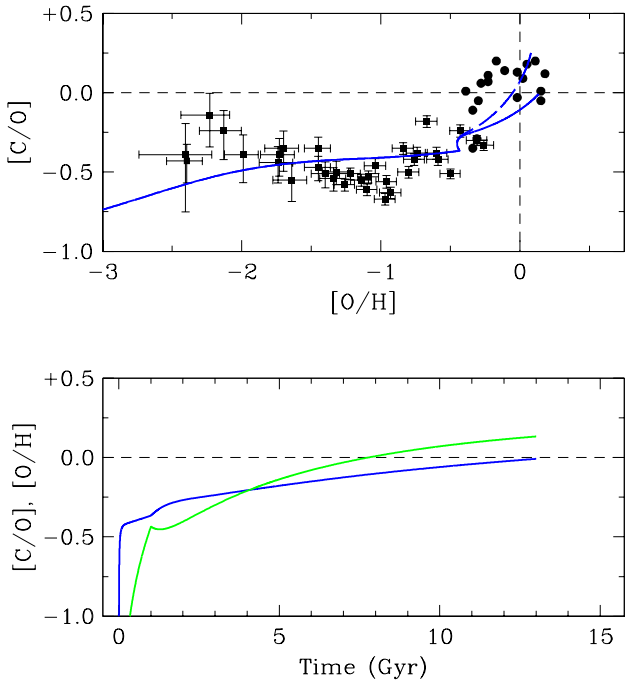}}
\vspace{-3.55cm}
\caption{The outputs of our chemical evolution
models are compared with the observed abundances of carbon
and oxygen (dots). In the upper panel, the continuous line shows
the behaviour of [C/O] vs. [O/H] produced by
our `standard' model, while the long-dash line shows the effect of
replacing the massive star yields of Meynet \& Maeder (2002) with those
of Maeder (1992). In the lower panel, the black line shows the 
rise in [C/O] with time in the `standard' model; 
the light grey line shows the same for [O/H].
}
\label{}
\end{figure}

\subsection{Outputs of chemical evolution models}

With the parameters described in Sections 5.1, 5.2, and 5.3 above,
and interpolating the yields linearly in metallicity and stellar
mass, we can follow the evolution of carbon and oxygen in the solar
vicinity from 13\,Gyr ago to the present. 
In principle, it is possible to optimise the agreement 
between models and observations by appropriately adjusting
some of the many ingredients of the models. However, we feel
that such an approach is of limited use, 
unless the changes involved are physically motivated. 
For example, it is obvious that
if one were free to make {\it ad hoc} adjustments to the IMF,
a better fit to the data may result. However, given the lack of
empirical evidence for variations in the IMF,
there is little justification for introducing this additional
variable. Thus, rather than striving to obtain the best
fit to the data, we are more interested in considering 
general {\it trends} in the behaviour of the carbon and oxygen 
abundances, and in what such trends can tell us about the
production of these two elements.

Figure 6 shows, as a function of [O/H] and time,  
the [C/O] ratio computed with our `standard' model 
which, to recap, uses the yields of Meynet \& Maeder
(2002) for massive stars, and those of van den Hoek \& Groenewegen (1997)
for low and intermediate mass stars.
During the first Gyr, which in our model corresponds to the
formation of the halo (Section 5.1), the oxygen abundance reaches
[O/H]\,$\simeq  -0.5$ and the [C/O] ratio grows quickly to a plateau
at [C/O]\,$\simeq -0.5$, a value which is in approximate agreement
with the ratio measured in most halo stars.
At the onset of disk formation (recognisable in Figure 6 from
the discontinuity in the model lines),
[C/O] grows over a period of 12\,Gyr by a factor of $\sim 3$
to reach solar proportions today.
The agreement between the model and the measurements 
in disk stars can be improved if we adopt the earlier 
work by Maeder (1992; long-dash line in Figure 6),
which used higher mass loss rates for massive stars 
(by factors of 2--3; see Meynet \& Maeder 2000)
and thus resulted in higher carbon yields than the later 
calculations by Meynet \& Maeder (2002). 
This gives an indication of the uncertainties
associated with theoretical yields (rather than implying that
stellar rotation should not be taken into account).

\begin{figure*}
\centerline{\resizebox{14.5cm}{!}{\includegraphics[angle=270.0]{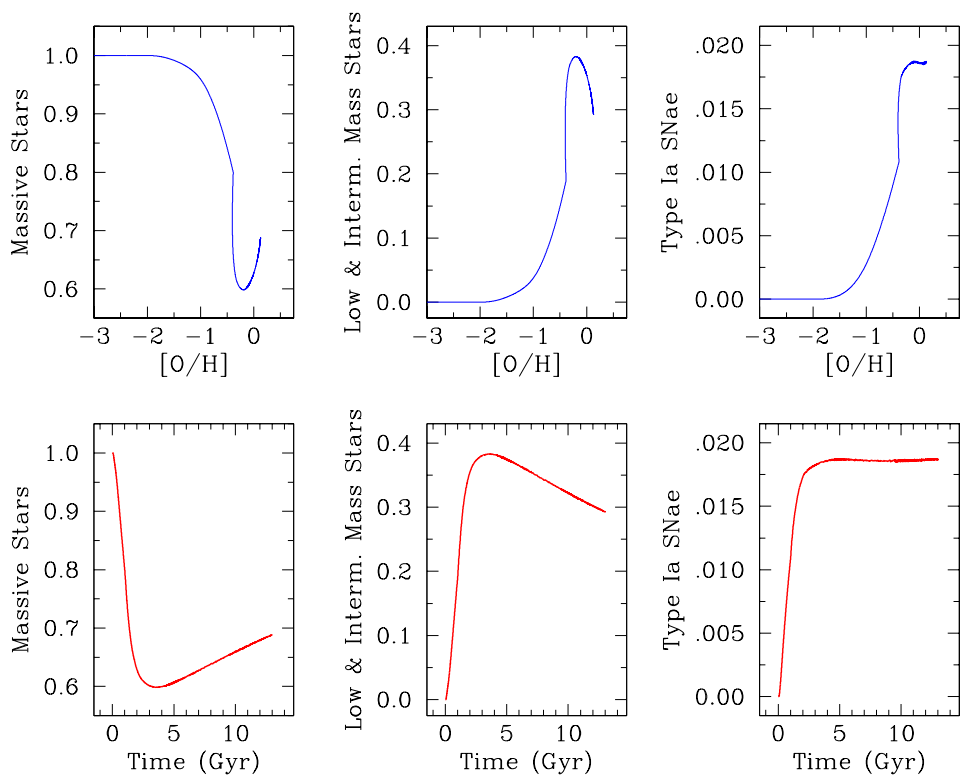}}}
\caption{The relative contributions to carbon enrichment from stars
of different masses in our models.
Left-hand panels: massive stars ($8\leq m \leq 80$\,$M_{\odot}$);
middle panels: low and intermediate mass stars ($0.8 \leq m \leq 8$\,$M_{\odot}$);
right-hand panels: Type Ia supernovae (4\% of stars with masses
in the range $3 \leq m \leq 16$\,$M_{\odot}$).
}
\label{}
\end{figure*}

The rise in [C/O] over the lifetime of the disk is due to
two effects, both working in the same direction. One is the
metallicity dependence of the carbon yield from massive stars.
At higher metallicity, massive stars undergo greater mass loss rates
with the net result that more carbon is ejected into the interstellar 
medium during the pre-supernova stage, before being converted into oxygen.
The second is the delayed contribution to the carbon yield by
low and intermediate mass stars, which take longer to evolve 
than massive stars. In our models, the former is always more 
important than the latter, as can be appreciated from inspection
of Figure 7. 

The question of whether massive stars, or stars of intermediate
and low mass, are the major sources of carbon is still
the subject of some debate. Our conclusion that the former
dominate is in line with those reached in similar studies by Henry et al. 
(2000), Prantzos (2003b), and Carigi (2003). It contrasts,
however, with the work of Chiappini et al. (2003a,b) who can reproduce
the rise of [C/O] in the disk entirely with 
intermediate mass stars by appropriately varying the
mass-loss parameter $\eta_{\rm AGB}$ in the yields by 
van den Hoek \&  Groenewegen (1997). 
Dray \& Tout (2003) have recently published theoretical models
of Wolf-Rayet stars over a range of metallicities and 
found comparable carbon enrichment from single WR stars and from asymptotic 
giant branch (AGB) stars. They concluded that whether the former or the latter are
the dominant source of carbon depends strongly on the set of 
AGB yields adopted and on the assumed IMF.

A dominant contribution from massive
stars, as illustrated in Figure 7, seems to us 
to be more consistent with the near-universal 
metallicity dependence of the [C/O] ratio,
as we now explain.
Essentially the same rise by a factor of $\sim 3$
exhibited by Milky Way stars when the metallicity
grows from $\sim 1/10$ solar to solar 
is also seen in \HII\ regions 
of nearby spiral and irregular galaxies
(see Figure 16 of Garnett 2003).
Even in the \lya\ forest at high redshift ($z \simeq 2 - 3$)
there is evidence for a substantial deficiency of carbon relative to oxygen
(Telfer et al. 2002).
In general, other spiral galaxies, irregular galaxies, 
and the sources responsible for seeding the \lya\ forest with metals 
will have experienced different star formation histories from
that of the Milky Way. Thus, the fact that the [C/O] ratio
shows approximately the same metallicity dependence in all of these
different environments favours a general explanation---such as 
the metallicity dependence of the carbon yields via massive star
winds---rather than an explanation based on time delay arguments
whose effects on the trend of [C/O] with [O/H] are contingent on 
the details of the previous history of star formation.

\subsection{Population III stars}

As explained above, the rise in the [C/O] ratio with metallicity
from the halo to the disk has been
known for several years and has been successfully modelled by
a number of previous studies 
(e.g. Henry et al. 2000;
Carigi 2003; Chiappini et al. 2003a,b; 
Prantzos 2003b).
{\it None} of these models, however, produces high values
of [C/O] at the lowest metallicities; in all of them
the general behaviour is similar to that of our `standard model'
in Figure 6, with [C/O] initially at very low levels and then
quickly increasing to a plateau at [C/O]\,$\simeq -0.5$.
As explained in Section 4, the new measurements presented in this paper
{\it may} indicate a different scenario, 
with [C/O] starting at near-solar values at the lowest
metallicities, and decreasing to [C/O]\,$\simeq -0.5$ as
[O/H] grows to 1/10 of solar. At the moment, 
the data provide no more than a hint that this may be the
case; the trend we see at low metallicities may be an artifact
of small-number statistics, or a reflection of metallicity-dependent
non-LTE corrections which we have not taken into account
in our derivation of the C/O ratios.
Nevertheless, it is instructive to consider possible explanations 
of this trend, should its reality be confirmed by future observations
of metal-poor halo stars and theoretical calculations of the \CI\ 
line formation.

\begin{figure}
\vspace{-1.05cm}
\hspace{-0.750cm}
\resizebox{10.3cm}{!}{\includegraphics{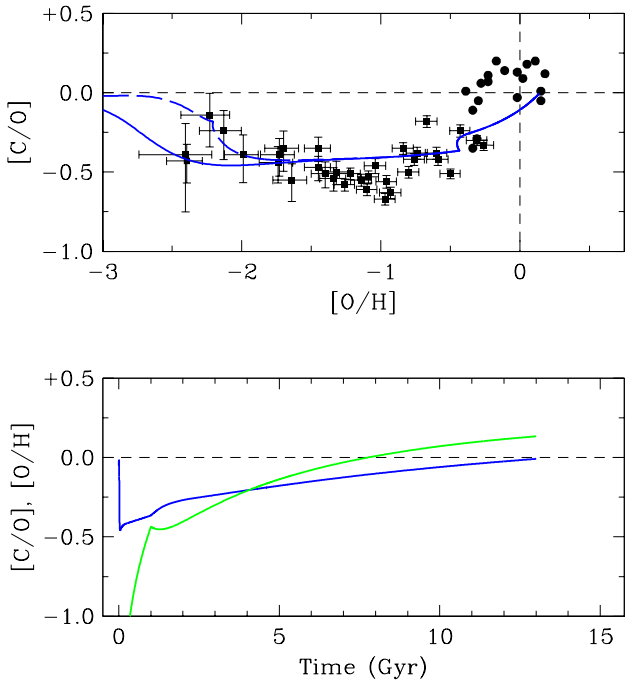}}
\vspace{-3.55cm}
\caption{The outputs of our chemical evolution
models obtained by adding to our `standard' model 
yields (applicable to stars with metallicities 
$Z \geq 10^{-5}$) the yields by Chieffi \& Limongi (2002)
for metal-free stars. Continuous lines: Population III
yields by Chieffi \& Limongi (2002) and normal 
(KTG) IMF. Long-dash line: same as the continuous line, 
but for a top-heavy IMF ($M \geq 10 M_{\odot}$).
In the lower panel, the black line shows the 
rise in [C/O] with time for the KTG IMF case;
the light grey line shows the same for [O/H].
}
\label{}
\end{figure}

Referring to Figure 6, it is evident that the sources 
of the relatively high carbon abundance 
must be associated with the first episodes
of star formation in the Milky Way, perhaps within the first few
hundred million years, when the metallicity was still very low
(recall that [O/H]\,$= -0.5$ is reached at the end 
of the halo formation, after just 1\,Gyr).
None of the published yields for stars with metallicities
$Z > 0$ can reproduce [C/O]\,$> -0.4$ when [O/H]\,$< -2$,
even allowing for (plausible) changes in the IMF.
We are thus drawn to consider zero metallicity stars.
Commonly referred to as Population III, such stars have attracted
considerable attention in recent years, mostly
for their cosmological importance as the first objects
to form in the universe and as the sources of reionisation
at $z > 6$. However, their masses and
chemical yields are still a matter of considerable speculation
(e.g. Woosley \& Weaver 1995; Heger \& Woosley 2002;
Chieffi \& Limongi 2002; 
Limongi \& Chieffi 2002; Umeda \& Nomoto 2002).

In Figure 8 we show the results of adding to our
`standard' model the C and O 
yields by Chieffi \& Limongi (2002)
for metallicities in the range $0 \leq Z  \leq 10^{-5}$.
Among the published Population~III yields these
are the only ones which 
produce the desired effect; in the combined
model shown in Figure~8
the first generation of stars enriches the gas
with carbon and oxygen in solar proportions and
the [C/O] ratio subsequently falls as
nucleosynthesis by Population~II stars
takes over. The nominal agreement with the observations
(given the uncertainties in the current limited dataset)
is improved if we assume that the IMF of Population~III
stars was top-heavy; as an example we show 
(long-dash line in Figure 8) a model
with the IMF truncated at 
$M_{\rm low} = 10 M_{\odot}$ 
(e.g. Hernandez \& Ferrara 2001;
Mackey, Bromm, \& Hernquist 2003;
Clarke \& Bromm 2003).

Chieffi \& Limongi (2002) have argued on different grounds
for a high carbon abundance at the end of He core burning
in metal-free stars and, by inference, favour the low
rate for the $^{12}$C\,($\alpha, \gamma$)\,$^{16}$O reaction
used in their set of nucleosynthesis calculations.
Possibly, the higher temperatures reached in the cores
of metal-free stars shift the balance between the 
$^{4}$He\,($2\alpha,\gamma$)\,$^{12}$C and
$^{12}$C\,($\alpha, \gamma$)\,$^{16}$O reactions
in favour of a higher carbon yield.
Or perhaps the processes responsible are related
to the mixing and fallback models 
proposed by Umeda \& Nomoto (2002, 2003) to explain
the abundance pattern of extremely metal-poor stars
with high energy supernova explosions
of massive Population III stars.
In any case it is clear that,
if further observations were to confirm that [C/O] really
was at near-solar values in the earliest stages
of the chemical evolution of the Milky Way, it would
be of great interest to investigate further
the physical reasons behind this effect as they
would provide a much needed window into 
the nucleosynthesis by the first generation of stars.

\section{Summary and Conclusions}

We have used UVES on the VLT to measure the abundances
of carbon and oxygen in 34 F and G dwarf and subgiant stars with 
halo kinematics and with metallicities in the range
from [Fe/H]\,$= -0.7$ to $-3.2$\,.
In our study we have targeted permitted,
high excitation
absorption lines of \CI\ near 9100\,\AA\ and 
of \OI\ near 7774\,\AA\ which are still
detectable (with equivalent widths of a few m\AA)
down to the lowest metallicities in our sample.
Line equivalent widths have been analysed with
1D LTE model atmospheres generated by the MARCS
code to deduce values of the C/O ratio. We show that
this ratio is probably insensitive to 3D effects because 
the lines are formed at similar 
deep levels within the stellar atmospheres.
However, we question the suggestion by Tomkin et al. (1992)
that corrections for departures from LTE are similar
for the two sets of lines and the C/O ratio is therefore
relatively insensitive to non-LTE effects. 
With more realistic estimates of the cross-sections
for inelastic hydrogen collisions, differential
non-LTE effects may be important (at the $\sim 0.2$\,dex
level) and metallicity dependent. Firm conclusions on this
important point await a full study of the structure of the 
\CI\ atom.

We consider our results together with those
of similar studies in disk stars to investigate
how the [C/O] ratio varies as a function 
of [O/H]. Carbon becomes proportionally less
abundant than oxygen as the oxygen abundance decreases
from solar; at [O/H]\,$\simeq -1$, [C/O]\,$\simeq -0.5$.
This metallicity dependence of the C/O ratio
is not confined to Galactic stars; a similar
drop in [C/O] with [O/H] has been revealed by
emission line studies of \HII\ regions in spiral
and irregular galaxies, and by analyses of C and
O absorption lines in the \lya\ forest at high
redshift. It thus appears to be a universal
effect which probably reflects the metallicity
dependence of the yields of carbon by massive stars
with mass loss. In the Milky Way, delayed release
of C by intermediate and low mass stars also contributes.
We can reproduce the behaviour of [C/O] vs. [O/H]
with a `standard' Galactic chemical evolution model,
and find that the relative contribution to carbon
enrichment from stars with masses $m > 8 M_{\odot}$
is $\simgt 60$\% throughout the lifetime
of the Galaxy.

Our survey also provides tentative evidence for 
an intriguing new trend which had not been 
recognised before: [C/O] may rise again in halo
stars with [O/H]\,$\simlt -1$. If real, such
an effect may indicate that the C/O ratio
started at near-solar levels in the earliest stages
of the chemical evolution of the Milky Way.
Among published work on the nucleosynthesis by
metal-free stars, the calculations by Chieffi \& Limongi (2002)
can reproduce the observed behaviour, particularly
if the IMF of Population III stars was top-heavy.
With the current limited statistics this is no
more than a $\sim 3 \sigma$ effect; it also remains
to be established to what extent it is affected by
systematic errors in the C/O ratios.
Thus, it is now a matter of priority to confirm, 
or refute, the reality of such a trend, both with
further observations of metal-poor halo stars
and sophisticated assessment of non-LTE effects.
Such tasks are well within current observational
and computational capabilities; we thus look forward
to a time in the relatively near future when 
the evolution of the C/O ratio in the Galaxy
will finally be clarified.

\begin{acknowledgements}
We are grateful to the ESO staff at Paranal for carrying out
the VLT/UVES service observations in their usual competent manner,
and to Francesca Primas and Vanessa Hill for their generous help
and advice with the preparation of the observing programme.
The interpretation of our results 
has benefited from discussions with several colleagues,
particularly Volker Bromm, Marco Limongi, and Chris Tout.
We thank Georges Meynet for communicating data on chemical
yields in advance of publication.
LC's work on this project is supported by
CONACyT grant 36904-E.
PEN acknowledges support for the Danish Natural Science Research Council (grant 
21-01-0523). MA has been supported by grants from the Swedish Natural Science
Research Council (grants F990/1999 and R521-880/2000), the Swedish 
Royal Academy of Sciences, the G\"oran Gustafsson Foundation and the
Australian Research Council (grant DP0342613).
\end{acknowledgements}

{}

\end{document}